\documentclass{emulateapj}
\usepackage{graphics}
\usepackage{psfig}
\usepackage{ulem}
\usepackage{longtable}

\def\sn{{SN~2006bp}}
\def\E{{\sl Einstein}}
\def\R{{\sl ROSAT}}
\def\A{{\sl ASCA}}
\def\C{{\sl Chandra}}
\def\B{{\sl BeppoSAX}}
\def\S{{\sl Swift}}
\def\X{{\sl XMM-Newton}}

\def\V{{\sl VLA}}

\def\gs{\mathrel{\mathchoice {\vcenter{\offinterlineskip\halign{\hfil
$\displaystyle##$\hfil\cr>\cr\sim\cr}}}
{\vcenter{\offinterlineskip\halign{\hfil$\textstyle##$\hfil\cr
>\cr\sim\cr}}}
{\vcenter{\offinterlineskip\halign{\hfil$\scriptstyle##$\hfil\cr
>\cr\sim\cr}}}
{\vcenter{\offinterlineskip\halign{\hfil$\scriptscriptstyle##$\hfil\cr
>\cr\sim\cr}}}}}

\begin{document}

\title{X-Ray, UV, and Optical Observations of Supernova 2006\lowercase{bp}
with Swift: \\ Detection of Early X-Ray Emission}

\author{S.~Immler\altaffilmark{1,2},
P. J. Brown\altaffilmark{3}, 
P. Milne\altaffilmark{4}, 
L. Dessart\altaffilmark{4}, 
P. A. Mazzali\altaffilmark{5,6},
W. Landsman\altaffilmark{1},
N. Gehrels\altaffilmark{7}, 
R. Petre\altaffilmark{1}, 
D. N. Burrows\altaffilmark{3}, 
J. A. Nousek\altaffilmark{3}, 
R. A. Chevalier\altaffilmark{8},
C. L. Williams\altaffilmark{1},
M. Koss\altaffilmark{1},
C. J. Stockdale\altaffilmark{9},
M. T. Kelley\altaffilmark{9},
K. W. Weiler\altaffilmark{10},
S. T. Holland\altaffilmark{1,2},
E. Pian\altaffilmark{5},
P. W. A. Roming\altaffilmark{3},
D. Pooley\altaffilmark{11,12},
K. Nomoto\altaffilmark{13},
J. Greiner\altaffilmark{14},
S. Campana\altaffilmark{15}, and
A. M. Soderberg\altaffilmark{16}
}

\email{stefan.m.immler@nasa.gov}
\altaffiltext{1}{Astrophysics Science Division, X-Ray Astrophysical Laboratory,
Code 662, NASA Goddard Space Flight Center, Greenbelt, MD 20771}
\altaffiltext{2}{Universities Space Research Association, 
10211 Wincopin Circle, Columbia, MD 21044}
\altaffiltext{3}{Department of Astronomy and Astrophysics, Pennsylvania 
State University, 525 Davey Laboratory, University Park, PA 16802}
\altaffiltext{4}{Steward Observatory, 933 North Cherry Avenue, RM N204 Tucson, 
AZ 85721}
\altaffiltext{5}{INAF, Osservatorio Astronomico di Trieste, via G.B. 
Tiepolo 11, 34131 Trieste, Italy}
\altaffiltext{6}{Max-Planck-Institut f\"ur Astrophysik, Karl-Schwarzschild
Strasse 1, 85741 Garching, Germany}
\altaffiltext{7}{Astrophysics Science Division, Astroparticle Physics Laboratory,
Code 661, NASA Goddard Space Flight Center, Greenbelt, MD 20771}
\altaffiltext{8}{Department of Astronomy, University of Virginia, 
P.O. Box 400325, Charlottesville, VA 22904}
\altaffiltext{9}{Department of Physics, Marquette University, 
P.O. Box 1881, Milwaukee, WI 53201-1881}
\altaffiltext{10}{Naval Research Laboratory, Code 7210, Washington, DC 20375-5320}
\altaffiltext{11}{Astronomy Department, University of California at Berkeley, 
601 Campbell Hall, Berkeley, CA 9472}
\altaffiltext{12}{Chandra Fellow}
\altaffiltext{13}{Department of Astronomy, School of Science, University of 
Tokyo, Bunkyo-ku, Tokyo 113-0033, Japan}
\altaffiltext{14}{Max-Planck-Institut f\"ur extraterrestrische Physik, 
Giessenbachstrasse, 85748 Garching, Germany}
\altaffiltext{15}{INAF - Osservatorio Astronomico di Brera, via E. Bianchi 46, I-23807
Merate, Italy}
\altaffiltext{16}{Division of Physics, Mathematics, and Astronomy, 
California Institute of Technology, MS 105-24, Pasadena, CA 91125}

\shorttitle{X-Ray, UV, and Optical Observations of SN 2006bp}
\shortauthors{Immler et~al.}

\begin{abstract}
We present results on the X-ray and optical/UV emission from the type~IIP
supernova (SN) 2006bp and the interaction of the SN shock with its environment,
obtained with the X-Ray Telescope (XRT) and UV/Optical Telescope (UVOT) 
on-board the \S\ observatory. SN~2006bp is detected in X-rays 
at a $4.5\sigma$ level of significance in the merged XRT data from days 1 to 
12 after the explosion. If the (0.2--10~keV band) X-ray 
luminosity of $L_{0.2-10}=(1.8\pm0.4) \times10^{39}~{\rm ergs~s}^{-1}$ is caused by 
interaction of the SN shock with circumstellar material (CSM), deposited by a 
stellar wind from the progenitor's companion star, a mass-loss rate of
$\dot{M} \approx 1 \times 10^{-5}~M_{\odot}~{\rm yr}^{-1}~(v_{\rm w}/10~{\rm km~s}^{-1})$
is inferred. The mass-loss rate is consistent with the non-detection in the radio 
with the \V\ on days 2, 9, and 11 after the explosion and characteristic of a red 
supergiant progenitor with a mass around $\approx 12$--$15~{\rm M_\odot}$ prior 
to the explosion. The \S\ data further show a fading of the X-ray emission starting 
around day 12 after the explosion. 
In combination with a follow-up \X\ observation obtained on day 21 after the 
explosion, an X-ray rate of decline $L_{\rm x} \propto t^{-n}$ with index 
$n=1.2\pm0.6$ is inferred. Since no other SN has been detected in X-rays 
prior to the optical peak and since type~IIP SNe have an extended 'plateau'
phase in the optical, we discuss the scenario that the X-rays might be 
due to inverse Compton scattering of photospheric optical photons off 
relativistic electrons produced in circumstellar shocks. However, due to the high 
required value of the Lorentz factor ($\approx 10$--$100$), inconsistent with
the ejecta velocity inferred from optical line widths, we conclude that Inverse 
Compton scatterring is an unlikely explanation for the observed X-ray emission.
The fast evolution of the optical/ultraviolet (1900--5500\AA) spectral energy 
distribution and the spectral changes observed with \S\ reveal the onset of 
metal line-blanketing and cooling of the expanding photosphere during the first few 
weeks after the outburst.
\end{abstract}

\keywords{stars: supernovae: individual (SN 2006bp) --- 
stars: circumstellar matter ---
X-rays: general --- 
X-rays: individual (SN 2006bp) --- 
X-rays: ISM --- 
ultraviolet: ISM}

\section{Introduction}
\label{introduction}

\sn\ was discovered on April 9.6, 2006, with an apparent magnitude 
of 16.7 in unfiltered 20-s CCD exposures using a 0.60-m f/5.7 reflector
(Nakano \& Itagaki 2006). Subsequent observations by Itagaki showed that the 
SN brightened rapidly between April 9.6 and 9.8 UT. The non-detection in 
unfiltered ROTSE-IIIb observations on April 9.15 UT ($>16.9$~mag; Quimby 
et~al.\ 2006) give a likely explosion date of April 9, 2006. 
Based on the optical/UV colors (Immler, Brown \& Milne 2006)
observed with \S\ and the detection of hydrogen and helium lines in ground-based
optical spectra (Quimby et~al.\ 2006), \sn\ was classified as a core-collapse 
type~II SN.

The optical/UV lightcurve established by \S\ (this work) shows that \sn\ belongs 
to the class of type~IIP SNe which are characterized by a prolonged 'plateau' 
period of sustained optical flux (see Sec.~\ref{uvotresults}).
This plateau phase results from recombination in the massive hydrogen envelopes 
($\gs2M_{\odot}$ H mass) of the progenitors. 
Most massive stars ($\approx 8$--$25~M_{\odot}$) become type II SNe, approximately 
2/3 of which are type IIP (Heger et~al.\ 2003) depending on the mass and
metallicity. Due to the large masses of the progenitor stars and high abundance
of type IIP SNe, they play a primary role in the formation of neutron stars.
Such massive progenitor stars have strong stellar winds which can deposit 
significant amounts of material in their environments. 
As the outgoing SN shock runs through the CSM deposited 
by the stellar wind, large amounts of X-ray emission and radio synchrotron
emission can be produced. Four of the 26 SNe detected in X-rays to date\footnote{see
http://lheawww.gsfc.nasa.gov/users/immler/supernovae\_list.html
for a complete list of X-ray SNe and references} are type IIP: SNe 1999em, 1999gi,
2002hh, and 2004dj (see Tables~1, 2 and references in Chevalier et~al.\ 2006).
Mass-loss rate estimates from the X-ray and radio observations are between a few 
$\times 10^{-6}$ to $10^{-5}~M_{\odot}~{\rm yr}^{-1}~(v_{\rm w}/10~{\rm km~s}^{-1})$,
where $v_{\rm w}$ is the pre-SN stellar wind velocity.

In this paper we present the earliest observation and detection of a SN in X-rays
to date, starting one day after the explosion. In addition to the \S\
X-ray and a follow-up \X\ X-ray observation, we discuss the broad-band spectral
energy distribution during the early evolution (days to weeks) of \sn, as well
as multi-epoch UV spectra obtained with \S. 


\begin{figure*}[t!]
\unitlength1.0cm
   \begin{picture}(6,6) 
\put(-0,0){ \begin{picture}(6,6)
	\psfig{figure=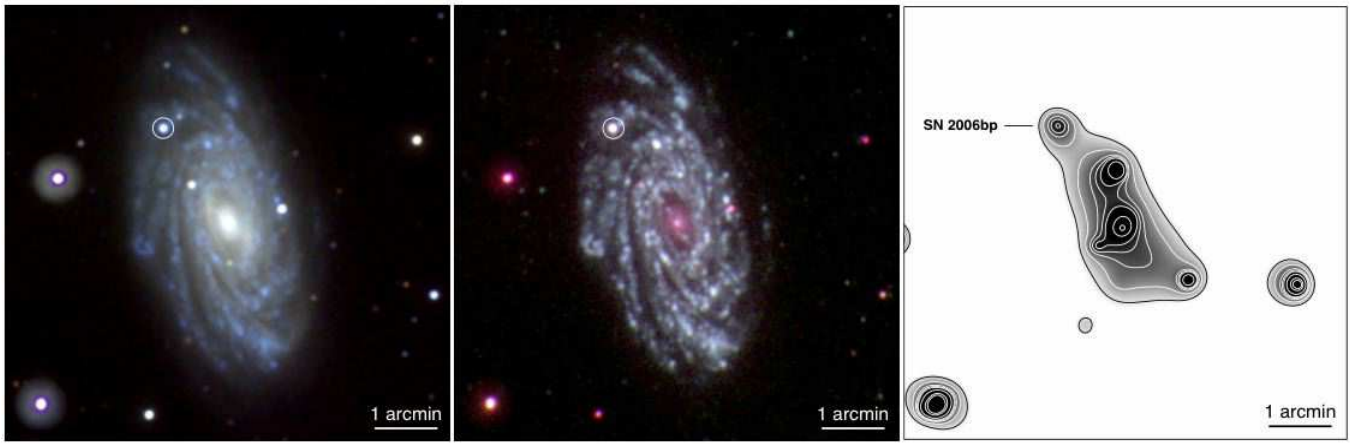,width=18cm,clip=}
	\end{picture}
	}
    \end{picture}
\caption{
{\sl Swift} optical, UV, and X-ray image of \sn\ and its host galaxy NGC~3953. 
{\bf Left-hand panel:}
\S\ optical image, constructed from the UVOT $V$ (750 s exposure time; red), 
$B$ (1,966 s; green) and $U$ (1,123 s; blue) filters obtained from 34 co-added
images taken between 2006-04-10.54 UT and 2006-05-30.45 UT. The optical images
are slightly smoothed with a Gaussian filter of 1.5 pixel (FWHM). The position 
of \sn\ is indicated by a white circle of 10 arcsec radius. The size of the 
image is $6.5~{\rm arcmin} \times 6.5~{\rm arcmin}$. 
{\bf Middle panel:}
The UV image was constructed from the co-added UVOT $UVW1$ (750 s exposure time; 
red), $UVM2$ (1,966 s; green) and $UVW2$ (1,123 s; blue) filters obtained 
between 2006-04-10.54 UT and 2006-05-30.45 UT. The UV images are 
slightly smoothed with a Gaussian filter of 1.5 pixel (FWHM).
Same scale as the left-hand panel.
{\bf Right-hand panel:} The (0.2--10 keV) X-ray image was constructed from 
the merged 41.4~ks XRT data and is adaptively smoothed to archive a S/N in
the range 2.5--4. Contour levels are 0.3, 0.6, 0.9, 1.5, 3, 6, 12, and 
20 counts pixel$^{-1}$. Same scale as the optical and UV images.
} 
\label{fig1}
\end{figure*}

\section{Observations}
\label{obs}

\subsection{{\sl Swift} UVOT Optical/UV Observations}
\label{uvot}

The Ultraviolet/Optical Telescope (UVOT; Roming et~al.\ 2005) and X-Ray Telescope
(XRT, Burrows et~al.\ 2005) on-board the \S\ Observatory (Gehrels et~al.\ 2004) 
began observing \sn\ on April 10.54 UT. 
The HEASOFT\footnote{http://heasarc.gsfc.nasa.gov/docs/software/lheasoft/} 
(version 6.1.1) and \S\ Software (version 2.5, build 19) tools and latest 
calibration products were used to analyze the data.

The SN was detected with the UVOT at ${\rm R.A.} = 11{\rm h}53{\rm m}55{\rm s}.70$, 
${\rm Decl.} = +52^{\rm o}21'10\farcs4$ (equinox 2000.0), with peak magnitudes of
$V = 15.1$, $B = 15.4$, $U = 14.5$, $UVW1$ [216--286~nm FWHM] $= 14.7$, 
$UVM2$ [192--242~nm] $= 15.1$, and $UVW2$ [170--226~nm] $= 14.8$. 
Statistical and systematic errors are 0.1~mag each.

Based on the UVOT photometry and $V-B$ and $B-U$ colors, the SN was 
classified as a young type~II event (Immler, Brown \& Milne 2006).
The classification was confirmed by Hobby-Eberly Telescope (HET) spectra 
taken on day 2, showing a blue continuum and a narrow absorption line at 592~nm, 
consistent with Na~I in the host galaxy NGC~3953 rest frame ($z = 0.00351$, 
Verheijen \& Sancisi 2001), a narrow emission line consistent with rest-frame 
H$\alpha$, a broad absorption line around 627~nm, and a narrow but slightly 
broadened emission line at 583~nm (Quimby et~al.\ 2006).

Thirty-five individual exposures were obtained between April 10.54 UT 
and May 30.45 UT (see Table~1 for an observation log). Assuming an explosion 
date of April 9, the observations correspond to days 1--51 after the explosion.

\S\ optical, UV, and X-ray images of \sn\ and its host galaxy NGC~3953 
(Hubble type SBbc; NED) are shown in Fig.~1. The UVOT 6-filter lightcurve of
\sn\ is given in Fig.~2.

\begin{deluxetable}{cccc}
\tabletypesize{\footnotesize}
\tablecaption{{\sl Swift} Observations of 2006\lowercase {bp} \label{tab1}}
\tablewidth{0pt}
\tablehead{
\colhead{Sequence} &
\colhead{Date} &
\colhead{XRT Exp} &
\colhead{UVOT Exp} \\
\colhead{} &
\colhead{[UT]} &
\colhead{[s]} &
\colhead{[s]} \\
\noalign{\smallskip}
\colhead{(1)}  &
\colhead{(2)}  &
\colhead{(3)}  &
\colhead{(4)}}
\startdata
30390001 &  2006-04-10~12:47:26 & 4333 & 4917 \\
30390003 &  2006-04-11~03:22:00 & 4323 & 4301 \\
30390004 &  2006-04-12~08:19:01 & 3157 & 2770 \\
30390005 &  2006-04-12~03:30:00 & 1705 & 1684 \\
30390006 &  2006-04-13~03:37:00 & 6906 & 6660 \\
30390007 &  2006-04-14~08:38:01 & \phantom{0}748 & \phantom{0}746 \\
30390008 &  2006-04-14~16:39:01 & \phantom{0}868 & \phantom{0}899 \\
30390009 &  2006-04-14~08:34:01 & \phantom{0}156 & \phantom{0}155 \\
30390010 &  2006-04-14~16:36:01 & \phantom{0}\phantom{0}\phantom{0}4 & \phantom{0}\phantom{0}\phantom{0}4 \\
30390011 &  2006-04-15~07:06:01 & 3373 & 3296 \\
30390012 &  2006-04-15~18:22:01 & 1572 & 1544 \\
30390013 &  2006-04-16~05:36:00 & \phantom{0}260 & \phantom{0}260 \\
30390014 &  2006-04-16~18:29:00 & \phantom{0}\phantom{0}80 & \phantom{0}\phantom{0}81 \\
30390015 &  2006-04-16~05:39:01 & 3013 & 2995 \\
30390016 &  2006-04-16~18:30:02 & 1537 & 1526 \\
30390017 &  2006-04-17~04:04:01 & 3666 & 3494 \\
30390018 &  2006-04-17~18:35:01 & 1639 & 1612 \\
30390019 &  2006-04-18~07:23:00 & \phantom{0}\phantom{0}61 & \phantom{0}\phantom{0}57 \\
30390020 &  2006-04-18~07:27:01 & 2609 & 2935 \\
30390021 &  2006-04-20~14:00:01 & 3174 & 3118 \\
30390022 &  2006-04-21~06:04:01 & \phantom{0}363 & \phantom{0}364 \\
30390023 &  2006-04-21~06:09:01 & 2741 & 2743 \\
30390024 &  2006-04-22~02:57:00 & 2801 & 2749 \\
30390025 &  2006-04-23~03:03:01 & \phantom{0}742 & \phantom{0}743 \\
30390026 &  2006-04-23~03:07:01 & 5248 & 5221 \\
30390027 &  2006-04-24~01:32:01 & 2999 & 2733 \\
30390028 &  2006-04-26~01:43:00 & 1782 & 1642 \\
30390029 &  2006-04-28~00:50:01 & 2829 & 2693 \\
30390030 &  2006-05-01~10:14:01 & \phantom{0}974 & \phantom{0}965 \\
30390031 &  2006-05-01~11:50:01 & 3793 & 3914 \\
30390032 &  2006-05-04~04:06:01 & 2542 & 2450 \\
30390034 &  2006-05-10~00:21:40 & \phantom{0}884 & \phantom{0}873 \\
30390036 &  2006-05-22~12:17:00 & \phantom{0}622 & \phantom{0}863 \\
30390037 &  2006-05-22~12:34:01 & \phantom{0}633 & \phantom{0}976 \\
30390038 &  2006-05-30~08:32:01 & 5242 & 5143
\enddata
\tablecomments{
(1)~{\sl Swift} sequence number;
(2)~Date in units of Universal Time (UT);
(3)~X-Ray Telescope (XRT) exposure time in units of s;
(4)~Ultraviolet/Optical Telescope (UVOT) exposure time in units of s.
}
\end{deluxetable}

\subsection{{\sl Swift} Grism Observations}
\label{grism}

The UVOT includes two grisms for obtaining spectra in the UV
and visible bands. The UV grism has a nominal wavelength range of
1800---2900~\AA, while the V grism has a wavelength range of 2800--5200~\AA.
The UV grism also records spectra in the 2900--4900~\AA\ range, but at half
the sensitivity of the V grism, and with the possibility of
contamination by second order overlap.  We obtained six UV grism
spectra and one V grism spectrum of \sn. The four earlier grism
observations had moderate to severe contamination from the host galaxy
and bright stars in the field and are not used in this study.
The successful grism observations are shown in Fig.~3. A log of the observations
is given in Table~2.

\begin{deluxetable}{cccccc}
\tabletypesize{\footnotesize}
\tablecaption{UVOT Grism Observations of 2006\lowercase {bp} \label{tab2}}
\tablewidth{0pt}
\tablehead{
\colhead{$t$} &
\colhead{Sequence} &
\colhead{Mode} &
\colhead{$T_{\rm start}$} &
\colhead{Exp} &
\colhead{$N_{\rm exp}$} \\
\noalign{\smallskip}
\colhead{[d]} &
\colhead{} &
\colhead{} &
\colhead{[UT]} &
\colhead{[s]} &
\colhead{} \\
\noalign{\smallskip}
\colhead{(1)}  &
\colhead{(2)}  &
\colhead{(3)}  &
\colhead{(4)}  &
\colhead{(5)}  &
\colhead{(6)}
}
\startdata
\phantom{0}9 & 00030390020 &  UV & 2006-04-18T07:30:28 &  2,431  &  2 \\
12	     & 00030390023 &  \phantom{0}V & 2006-04-21T06:10:22 &  2,688  &  2 \\
14	     & 00030390026 &  UV & 2006-04-23T03:10:13 &  3252  &  4
\enddata
\tablecomments{
(1)~Days after the explosion of \sn\ (April 9, 2006);
(2)~{\sl Swift} sequence number;
(3)~UVOT grism mode;
(4)~Exposure start time in unites of UT;
(5)~Exposure time in units of s;
(6)~Number of individual exposures.
}
\end{deluxetable}

\subsection{{\sl Swift} XRT Observations}
\label{xrt}

The \S\ XRT observations were obtained simultaneously with the UVOT and grism 
observations. X-ray counts were extracted from a circular region with an aperture 
of 10~pixel ($24''$) radius centered at the position of the SN. The background 
was extracted locally from a source-free region of radius of $1'$ to account 
for detector and sky background, and for a low level of residual diffuse 
emission from the host galaxy. 

\subsection{{\sl XMM-Newton} Observation}
\label{xmm}

An \X\ European Photon Imaging Camera (EPIC) Director's Discretionary Time
(DDT) observation was performed on April 30.39 UT (PI Immler, OBS-ID 0311791401), 
corresponding to day 21 after the explosion. 
SAS\footnote{http://xmm.vilspa.esa.es/sas/} (version 7.0.0),
FTOOLS\footnote{http://heasarc.gsfc.nasa.gov/docs/software.html} (version 6.1.1), 
and the latest \X\ calibration products were used to analyze the \X\ data.
Inspection of the EPIC PN and MOS data for periods with a high particle background 
showed no contamination of the data, which resulted in clean exposure times of 
21.2~ks for the PN and 22.9~ks for each of the two MOS instruments.

\subsection{{\sl VLA} Radio Observation}
\label{vla}
Radio observations were performed with the {\sl VLA}\footnote{The {\sl VLA} 
telescope of the National Radio Astronomy Observatory is operated by 
Associated Universities, Inc. under a cooperative agreement with the 
National Science Foundation} at three epochs (days 2, 9, and 11) with a 
search for radio emission conducted 
within a radius of $10''$ of the optical position (Nakano \& Itagaki, 2006) 
using AIPS\footnote{http://www.aoc.nrao.edu/aips/}.
 
Upper limits ($3\sigma$) to any radio flux density were established
at $< 0.414$~mJy on 2006 April 11.97 UT (spectral luminosity 
$<1.1 \times 10^{26}~{\rm erg s}^{-1}~{\rm Hz}^{-1}$ for an assumed distance 
of 14.9~Mpc) at 22.46~GHz (wavelength~1.3 cm) and $< 0.281$~mJy at 8.460~GHz 
(wavelength 3.5~cm); on 2006 April 18.27 at $< 0.498$~mJy at 14.94~GHz 
(wavelength 2.0~cm); and on 2006 April 20.16 at $< 0.276$~mJy at 22.46~GHz 
and $< 0.143$~mJy at 8.460~GHz (Kelley et~al.\ 2006).
Although only uncertain mass-loss rate estimates can be obtained from radio
non-detections, under standard assumptions ($v_{\rm w}=10~{\rm km~s^{-1}}$; 
$v_{\rm s}=10^4~{\rm km~s^{-1}}$; $T_{\rm CSM}=3$--$10 \times 10^4$~K; $m=1.0$;
see, e.g. Weiler et~al.\ 2002 and Lundqvist \& Fransson 1988), the upper limit
to the radio-determined pre-SN mass-loss rate is estimated to be 
$10^{-6}$ to $10^{-5}~M_{\odot}~{\rm yr}^{-1}$.


%

\section{Results}
\label{results}

\begin{figure}[t!]
\unitlength1.0cm
    \begin{picture}(6.5,6.5) 
\put(-0.2,0){ \begin{picture}(6.5,6.5)
	\psfig{figure=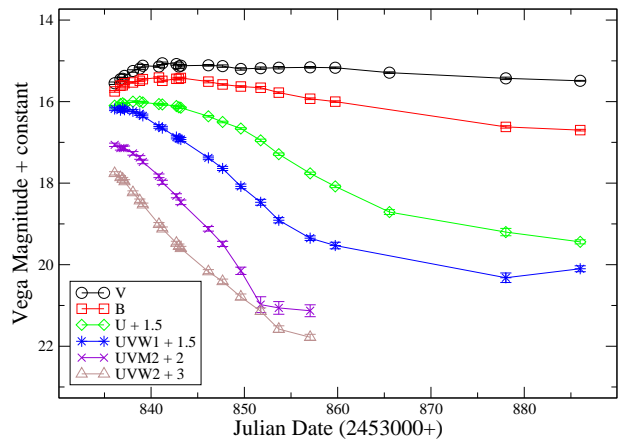,width=9.6cm,angle=-90,clip=}
	\end{picture}
	}
    \end{picture}
\caption{
{\sl Swift} UVOT lightcurve of \sn\ obtained in all six filters.
The $U$, $UVW1$, $UVM2$, and $UVW2$ lightcurves were shifted down vertically
by 1.5, 1.5, 2, and 3~mag, respectively, to avoid overlap.
} 
\label{fig2}
\end{figure}

\subsection{{\sl Swift} UVOT Optical/UV Observations}
\label{uvotresults}

The \S\ $V$-band UVOT observations show that the SN reached its maximum 
brightness around day seven after the estimated time of explosion, faded 
slightly over the next few days and subsequently reached a plateau phase 
characteristic for type~IIP SNe (see Fig.~2). 

At increasingly shorter wavelengths, the SN shows an earlier maximum and steeper 
rates of decline. In the roughly linear period 5--15 days after the explosion, 
the decay slopes are approximately 0.21, 0.25, 0.16, 0.06, 0.02, and 0.01 mags/day
in the $UVW2$, $UVM2$, $UVW1$, $U$, $B$, and $V$ bands, respectively. \\ \\

\subsection{{\sl Swift} and \X\ X-ray Observations}

An X-ray source is detected in the merged 41.4~ks \S\ XRT observations 
obtained between days 1--12 after the explosion at 
${\rm R.A.} = 11{\rm h}53{\rm m}56{\rm s}.1$, 
${\rm Decl.} = +52^{\rm o}21'11\farcs1$ (positional error $3\farcs5$), 
consistent with the optical position of the SN. The aperture-, background- 
and vignetting-corrected net count rate is 
$(1.3\pm0.3) \times 10^{-3}~{\rm cts~s}^{-1}$ (0.2--10 keV band).
Two additional X-ray sources are detected within the $D_{25}$ diameter of
the host galaxy, as well as an X-ray source associated with the nucleus of the
host galaxy (see Fig.~1). The chance probability of any of the three X-ray sources 
being within a radius of $3\farcs5$ at the position of \sn\ is estimated to be 
$1.5 \times 10^{-3}$.
 
Adopting a thermal plasma spectrum with a temperature of $kT = 10$~keV 
(see Fransson, Lundqvist \& Chevalier 1996 and therein) and assuming a Galactic 
foreground column density with no intrinsic absorption 
($N_{\rm H} = 1.58 \times 10^{20}~{\rm cm}^{-2}$; Dickey \& Lockman 1990)
gives a 0.2--10 keV X-ray band unabsorbed flux and luminosity of 
$f_{0.2-10} = (6.8 \pm 1.6) \times 10^{-14}~{\rm ergs~cm}^{-2}~{\rm s}^{-1}$ and 
$L_{0.2-10} = (1.8 \pm 0.4) \times 10^{39}~{\rm ergs~s}^{-1}$, 
respectively, for a distance of 14.9~Mpc ($z = 0.00351$, Verheijen \& Sancisi 2001; 
$H_0 = 71~{\rm km~s}^{-1}~{\rm Mpc}^{-1}$, $\Omega_{\Lambda}=2/3$, $\Omega_{\rm m}=1/3$).

The SN is not detected in merged 23.1~ks XRT data obtained between days 12--31
after the explosion. The ($3\sigma$) upper limit to the X-ray count rate is 
$<8.1 \times 10^{-4}~{\rm cts~s}^{-1}$ (0.2--10~keV), corresponding to
an unabsorbed flux and luminosity of 
$f_{0.2-10} < 4 \times 10^{-14}~{\rm ergs~cm}^{-2}~{\rm s}^{-1}$ and 
$L_{0.2-10} < 1 \times 10^{39}~{\rm ergs~s}^{-1}$.

In order to probe the evolution of the X-ray emission in more detail, we subdivided 
the data taken before day 12 into two observations with similar exposure times
of $\approx 20$~ks each (see Table~3). Within the errors of the photon statistics,
no significant decline is observed over the first 12 days after the explosion
(see Fig.~4).

To follow the X-ray rate of decline over a longer period, an \X\
observation was obtained on day 21 after the explosion. The SN is detected
at a $4.9 \sigma$ level of confidence, with an EPIC PN net count rate of
$(3.0\pm0.6) \times 10^{-3}~{\rm cts~s}^{-1}$ (0.2--10 keV band),
corresponding to an unabsorbed flux and luminosity of 
$f_{0.2-10} = (1.4\pm0.3) \times 10^{-14}~{\rm ergs~cm}^{-2}~{\rm s}^{-1}$ and 
$L_{0.2-10} = (3.8\pm0.8) \times 10^{38}~{\rm ergs~s}^{-1}$.
A best fit X-ray rate of decline of $L_{\rm x} \propto t^{-n}$ with index 
$n=1.2\pm0.6$ is obtained using the \S\ XRT and \X\ detections.
The X-ray lightcurve of \sn\ is given in Fig.~4. Due to the limited photon 
statistics of the \S\ XRT and \X\ EPIC data, no detailed spectral fitting 
is possible.

\begin{figure}[t!]
\unitlength1.0cm
    \begin{picture}(6,6)
\put(-0.5,0){ \begin{picture}(6,6)
	\psfig{figure=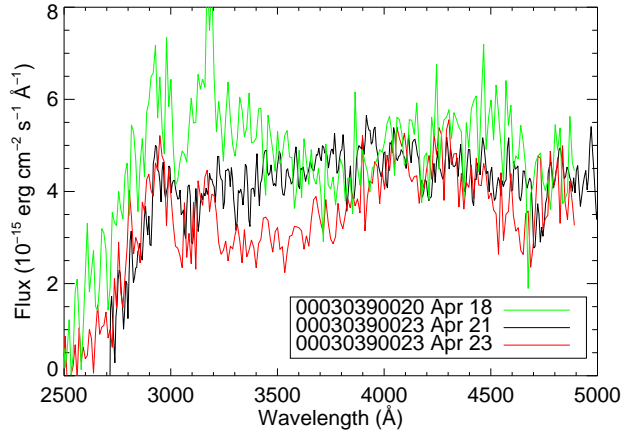,width=9cm,clip=}
	\end{picture}
	}
    \end{picture}
\caption{
{\sl Swift} UV grism observations of \sn\ obtained on April 18, 21, and 23, 
2006, corresponding to days 9, 12, and 14 after the explosion.
} 
\label{fig3}
\end{figure}

\begin{deluxetable*}{ccccccccccc}
\tabletypesize{\footnotesize}
\tablecaption{X-Ray Observations of 2006\lowercase{bp} \label{tab3}}
\tablewidth{0pt}
\tablehead{
\colhead{Day after} &
\colhead{Instrument} &
\colhead{Exposure} &
\colhead{S/N} &
\colhead{Rate} &
\colhead{$f_{\rm x}$} &
\colhead{$L_{\rm x}$} \\
\colhead{Explosion} &
\colhead{} &
\colhead{[ks]} &
\colhead{[$\sigma$]} &
\colhead{[$10^{-3}$]} &
\colhead{[$10^{-14}$]} &
\colhead{[$10^{39}$]} \\
\noalign{\smallskip}
\colhead{(1)}  &
\colhead{(2)}  &
\colhead{(3)}  &
\colhead{(4)}  &
\colhead{(5)}  &
\colhead{(6)}  &
\colhead{(7)}
}
\startdata
1--5 	& \S\ XRT  & 20.0 & 4.4  & $1.31\pm0.30$ & $6.94\pm1.58$ & $1.84\pm0.42$ \\
6--12  	& \S\ XRT  & 21.5 & 4.4  & $1.25\pm0.29$ & $6.58\pm1.52$ & $1.75\pm0.40$ \\
13--31	& \S\ XRT  & 23.1 & $<3$ & $<0.81$ & $<4.00$ & $<1.06$ \\
21	& {\sl XMM} EPIC PN & 21.2 & 4.9  & $3.0\pm0.6$ & $1.4\pm0.3$ & $0.38\pm0.08$
\enddata
\tablecomments{
(1)~Days after the explosion of \sn\ (April 9, 2006);
(2)~Instrument used;
(3)~Exposure time in units of ks;
(4)~Significance of source detection in units of Gaussian sigma;
(5)~0.2--10~keV count rate in units of cts~ks$^{-1}$;
(6)~0.2--10~keV X-ray band unabsorbed flux in units of $10^{-14}~{\rm ergs~cm}^{-2}~{\rm s}^{-1}$;
(7)~0.2--10~keV X-ray band luminosity in units of $10^{39}~{\rm ergs~s}^{-1}$.
}
\end{deluxetable*}

\section{Discussion}
\label{discussion}

\subsection{Ultraviolet Emission}
\label{uv}

UV emission from SNe explosions can arise from different physical 
processes. The first photon signature from a core-collapse SN event is thought 
to be associated with shock break-out, which should manifest itself in a burst 
of light peaking in the X-rays and the far-UV (Ensman \& Burrows 1992; 
Blinnikov \& Bartunov 1993; Blinnikov et~al.\ 1998, 2000; Li 2006).
The exact duration of this emission can be from hours to days, depending on the 
progenitor structure. Fast expansion of the ejecta and efficient radiative cooling 
at its 'photosphere' make this epoch short lived and therefore difficult to observe.  
At such early times, the object is not bright in the optical, and is thus usually 
still undetected. A SN is usually discovered when it becomes bright in the optical, 
but at this time the photosphere has cooled too much for X-ray emission.
Observational evidence for a shock break-out was provided by \S\ observations
of the Gamma-Ray Burst (GRB) related to SN~2006aj (Campana et~al.\ 2006), although
this was probably mediated by a cocoon of material lost by the star before it
collapsed.

UV emission has also been observed at early times ($\la$ 2 weeks) in the
photospheric phase of SN ejecta, emanating from the progenitor envelope layers 
that transition from thick to thin over time when the ejecta are still very hot
(e.g.\ Mazzali 2000). The UV emission is typically strongest during early times (as in the 
spectrum of SN~1987A on day one; Pun et~al.\ 1995) and fades substantially over the
course of a few days (weak UV emission has been detected in an HST observation of 
SN~1993J, 18 days after the explosion; Jeffery et~al.\ 1994; Baron et~al.\ 1994). 
At later times ($\gs$weeks), UV emission is in general associated 
with the interaction of the ejecta with the CSM (Fransson et~al.\ 1987, 2002; 
Pun et~al.\ 2002).
 
The UV is therefore a very sensitive spectral probe of the fast changing
conditions in the SN expanding photosphere during the earliest phases.

The first observation of \sn\ shows a spectral energy distribution (SED) peak in the UV, 
rather than the optical (Fig.~5), which supports that the SN was detection very early.
Thirty-five observations finely sample the lightcurve over 51 days after discovery, 
revealing a flux variation in the UV by two orders of magnitude. 
The redward shift of the SED stems from cooling of the photosphere. There is a direct 
effect in the reduction of the equivalent blackbody (or effective/electron) temperature
(e.g.\ Mazzali \& Chugai 1995), but the main effect is caused by the shift of the opacity
from species of higher to lower ionization, as discussed in Eastman \& Kirshner (1989) and
Dessart \& Hillier (2005, 2006).

The main physical process affecting the UV flux distribution is line blanketing (e.g.\ Lentz
et al. 2000, Mazzali 2000). Metal lines, especially lines of Fe{\sc ii} and Fe{\sc iii},
are very numerous in the UV, and the velocity dispersion in the ejecta causes these lines
to act as a blanket that blocks the UV flux. Photons absorbed in the UV are re-emitted at
visible wavelengths, where they can more easily escape (e.g.\ Mazzali \& Lucy 1993, Mazzali
2000). Thus the UV flux is determined by the ionization state of the SN envelope as well 
as by its metal content. Since on average UV Fe{\sc iii} lines are redder than Fe{\sc ii} 
lines, the sudden transition from Fe{\sc iii} to Fe{\sc ii} when hydrogen recombines in 
the ejecta of a type II SN shifts the region where line-blanketing is most effective and 
affects the UV spectrum dramatically (see Figs 1, 3, and 5 in Dessart \& Hillier 2005).
This onset of line blanketing is evident in the UV spectra of \sn\ as observed
by \S\ (see Fig.~3). The corresponding forest of overlapping lines gives an 
apparently continuous source of light blocking: at early times, Fe{\sc iii} blanketing
operates most strongly between 1500 and 2000\AA, while at later times, Fe{\sc ii} 
blanketing affects the entire UV range. Combined with this dominant background opacity 
are a few strong resonance lines of less abundant metals, e.g.\ Mg{\sc ii}~2800~\AA, 
which are only visible while metal line-blanketing is moderate.
   
The SED evolution supports the current understanding of the photospheric phase 
of type~II SNe, however, UV coverage will provide a more extended spectroscopic and 
photometric analysis. In particular, the unsaturated metal line blanketing together 
with the fast changing UV SED provides additional constraints on the 
temperature evolution, the ejecta ionization and composition, and, importantly, 
on reddening, that optical observations alone do not guarantee. 
This will be the subject of a forthcoming study (Dessart in preparation).

\subsection{X-Ray Emission}
\label{xray}

X-ray emission from young (days to weeks) SNe can be produced by the 
radioactive decay products of the ejecta, inverse Compton scattering of 
photospheric photons off relativistic electrons produced during the explosion,
as well as by the interaction of the SN shock with the ambient CSM (forward
shock) and SN ejecta (reverse shock).

While several of the emission lines characteristic of the radioactive 
decay products have been observed in SN~1987A (e.g., McCray 1993), 
the total X-ray output is five orders of magnitude lower 
($\approx 10^{34}~{\rm ergs~s}^{-1}$) than the observed X-ray luminosity 
of \sn. Therefore, no significant contribution of the radioactive decay 
products to the total X-ray luminosity of \sn\ is expected. 
Recent simulations of the expected X-ray emission from Compton-scattered 
$\gamma$-rays of the radioactive decay products of the SN ejecta for SN~2005ke
at early epochs (days) also shows the expected X-ray luminosity to be in
the range $10^{33}$--$10^{34}~{\rm ergs~s}^{-1}$ (Immler et~al. 2006), well
below the observed X-ray luminosity of \sn.

This leaves either inverse Compton scattering or CSM interaction as the
likely source of the detected X-ray emission.

Since type~IIP SNe have a prolonged plateau period with high optical output
associated with hydrogen recombination in the progenitor envelope, 
inverse Compton cooling of the
relativistic electrons produced during the explosion by the photospheric
photons might be important. Up-scattering of the optical photons to energies 
in the X-ray range (depending on the Lorentz factor and effective temperature 
of the photospheric emission) could produce a detectable X-ray flux during the 
first few days after outburst.

Chevalier, Fransson \& Nymark (2006) have shown that the X-ray luminosity of
inverse Compton emission takes the form
$\frac{dL_{\rm IC}}{dE} \approx 8.8\times 10^{38} \ \epsilon_r \gamma_{\rm min} 
\ E_{\rm keV}^{-1} \ \left(\frac{\dot M_{-6}}{v_{\rm w1}}\right) \ v_{s4} \ 
\left(\frac{L_{\rm bol}(t)}{10^{42}~{\rm ergs~s}^{-1}}\right) \ 
\left(\frac{t}{10~{\rm days}}\right)^{-1} \
\frac{\rm ergs}{\rm s~keV}$,
where $\epsilon_r$ is the fraction of the postshock energy density that is in 
relativistic electrons, $\gamma_{\rm min}$ the minimum Lorentz factor of 
the relativistic electrons, ${\dot M}_{-6}$ the mass-loss rate (in units of 
$10^{-6}~M_{\odot}~{\rm yr}^{-1}$), $v_{\rm w}$ the wind velocity (in units 
of ${\rm km~s}^{-1}$), $v_{s4}$ the SN shock velocity (in units of 
$10,000~{\rm km~s}^{-1}$), and $L_{\rm bol}(t)$ the bolometric luminosity at 
time $t$ after the explosion. Assuming a mass-loss rate of 
$\dot{M} = 2 \times 10^{-6}~M_{\odot}~{\rm yr}^{-1}~(v_{\rm w}/10~{\rm km~s}^{-1})$
(see below), a shock velocity of $15,000~{\rm km~s}^{-1}$, and a 
bolometric luminosity of $L_{\rm bol} = 10^{42}~{\rm ergs~s}^{-1}$, 
we estimate a monochromatic X-ray luminosity from inverse Compton scattering 
at 1~keV (near the peak of the \S\ XRT and \X\ EPIC response) of
$L_{\rm x} \approx \epsilon_r \gamma_{\rm min} \ 4 \times 10^{39}~{\rm ergs~s}^{-1}$.
The bolometric luminosity is justified by the modeling of SNe IIP lightcurves
(e.g., Chieffi et~al.\ 2003).


Since the value for $\epsilon_r$ is expected to be in the range
$\approx 0.01$--$0.1$, a Lorentz factor of $\gamma_{\rm min} \approx 10$--$100$ 
is needed to satisfy the condition that all of the observed X-rays are due to
inverse Compton scattering. As the most plausible value for the Lorentz factor 
for the shock velocities of SNe IIP is $\approx 1$ (Chevalier, Fransson \& Nymark 
2006), inverse Compton scattering is unlikely to account for the observed
X-ray emission of \sn, although it cannot be ruled out entirely.

Alternatively, the X-ray emission might be caused by the interaction of
the shock with the CSM, deposited by the progenitor's stellar wind.
Of all 26 SNe detected in X-rays over the past three decades, the importance 
of inverse Compton scattering has only been discussed for the type~Ic SNe~1998bw
(Pian et~al.\ 2000) and 2002ap (Bjornsson \& Fransson 2004), while the 
remaining X-ray SNe have been discussed in the context of thermal emission
[apart from the early emission of SN~1987A, see Park et~al.\ (2006) and references 
therein].

Assuming a constant mass loss rate $\dot{M}$ and wind velocity $v_{\rm w}$ 
from the progenitor's companion, the thermal X-ray luminosity of the forward 
shock region is
$L_{\rm x} = 1/(\pi m^2) \Lambda(T) \times (\dot{M}/v_{\rm w})^2\times(v_{\rm s} t)^{-1}$
(Immler et~al.\ 2006),
where $m$ is the mean mass per particle ($2.1\times10^{-24}$~g for a H+He 
plasma), $\Lambda (T)$ the cooling function of the heated plasma at temperature 
$T = 1.36 \times 10^{9} (n-3)^{2}/(n-2)^{2} (v_{\rm s}/[10^{4}~{\rm km/s}])^{2}$~K
(Chevalier \& Fransson 2003), $n$ is the ejecta density parameter (in the range 7--12) 
and $v_{\rm s}$ the shock velocity. Adopting an effective cooling function of 
$\Lambda_{\rm ff} = 2.4 \times 10^{-27}~g_{\rm ff} T_{\rm e}^{0.5}~{\rm ergs~cm}^3~{\rm s}^{-1}$
(Chevalier \& Fransson 2003) for an optically thin thermal plasma with a temperature 
of $T$ for the forward shock, where $g_{\rm ff}$ is the free-free Gaunt factor,
and $v_s = 15,000~{\rm km~s}^{-1}$, a mass-loss rate of 
$\dot{M} \approx 1 \times 10^{-5}~M_{\odot}~{\rm yr}^{-1}~(v_{\rm w}/10~{\rm km~s}^{-1})$ 
with an uncertainty of a factor of 2--3 is obtained.
Assuming different plasma temperatures in the range $10^7$--$10^9$ K would 
lead to changes in the emission measure by a factor of three. The X-ray 
luminosity from shock-heated plasma behind the reverse shock is assumed to be small 
compared to that of the forward shock since the expanding shell is still 
optically thick at such an early epoch. Since only high-quality X-ray 
spectra could give a conclusive answer to the contribution of the reverse 
shock to the total X-ray luminosity, our inferred mass-loss rate represents 
an upper limit.
In case absorption is small and the bulk of the thermal emission originates in the 
reverse shock, the value of $\dot{M}$ is reduced by a factor of up to 10 assuming an
ejecta density parameter of $n=10$ (with a range between 7 and 12; see equation 33 
in Chevalier 1982).

While other core-collapse SN (type~Ib/c and II) progenitors can produce mass-loss 
rates as high as $10^{-5}$ to $10^{-3}~M_{\odot}~{\rm yr}^{-1}$ due to the high masses 
and strong stellar winds of the progenitor stars, the mass-loss rate of the
\sn\ progenitor is similar to those of other type IIP SNe: a recent study of
the thermal X-ray and radio synchrotron emission of a sample of type IIP SNe 
(1999em, 1999gi, 2002hh, 2003gd, 2004dj, 2004et) gave mass-loss rate
estimates of a few $10^{-6}~M_{\odot}~{\rm yr}^{-1}$ (Chevalier, Fransson \&
Nymark 2006), characteristic of red supergiant progenitors, assuming stellar
wind velocities for red supergiants of 10--$15~{\rm km~s}^{-1}$. Comparison shows 
that the \sn\ progenitor had a mass-loss rate similar to the lowest in this sample
(SNe 1999gi, 2004dj) and an mass around $\approx 12$--$15~{\rm M_\odot}$ prior to
the explosion (see Fig.~1 in Chevalier, Fransson \& Nymark 2006) in case the X-ray
emission was dominated by the reverse shock.


CSM interaction gives an expected $t^{-1}$ decline of the X-ray flux, consistent 
with the observed rate of decline ($t^{-n}$ with index $n=1.2\pm0.6$). 
Inverse Compton scattering, on the other hand, only gives a $t^{-1}$ decline in 
case of $L_{\rm bol}(t)$ is constant over time (as in the UVOT $V$-band light curve).
However, at early times the blue/UV emission, which shows a fast rate of decline,
contributes significantly to the bolometric luminosity, which leads to a faster
X-ray rate of decline for Inverse Compton Scattering. Due to the large errors 
associated with the X-ray rate of decline, inverse Compton scattering cannot be 
entirely ruled out, but is a less likely scenario for the production of the 
observed X-rays.

\begin{figure}[t!]
\unitlength1.0cm
    \begin{picture}(6.2,6.2) 
\put(-0.5,0){ \begin{picture}(6.2,6.2)
	\psfig{figure=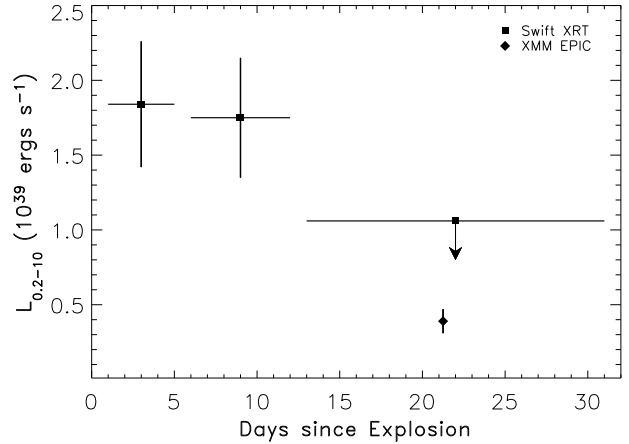,width=9.3cm,angle=0,clip=}
	\end{picture}
	}
    \end{picture}
\caption{
X-ray lightcurve (0.2--10~keV) of SN~2006bp as observed with the \S\ 
XRT and \X\ EPIC instruments. The time is given in days after the outburst 
(April 9, 2006). Vertical error bars are statistical $1\sigma$ errors; 
horizontal error bars indicate the periods covered by the observations 
(which are not contiguous). The \S\ XRT upper limit from day 13--31 is at 
a $3\sigma$ level of confidence.
} 
\label{fig3}
\end{figure}

The distinguishing characteristic of the two emission processes is their
spectrum (power law vs thermal plasma). In the absence of high-quality spectra 
for the \S\ XRT and \X\ EPIC data, such a distinction cannot be made since the 
\X\ EPIC data are equally well fit (using Cash statistics; Cash 1979) by a power law 
(best fit photon index $\Gamma = 1.6^{+1.2}_{-0.7}$; $\chi^2=42.6$, ${\rm d.o.f.}=52$), 
a thermal plasma spectrum (best fit temperature $kT = 1.7$--$27$~keV, consistent with 
our spectral assumptions to infer fluxes; $\chi^2=42.3$, ${\rm d.o.f.}=52$) and a 
thermal bremsstrahlung spectrum (best fit temperature $kT = 1.0$--$179$~keV; 
$\chi^2=41.4$, ${\rm d.o.f.}=52$).

\begin{figure}[t!]
\unitlength1.0cm
    \begin{picture}(6.5,6.5) 
\put(0,0){ \begin{picture}(6.5,6.5)
	\psfig{figure=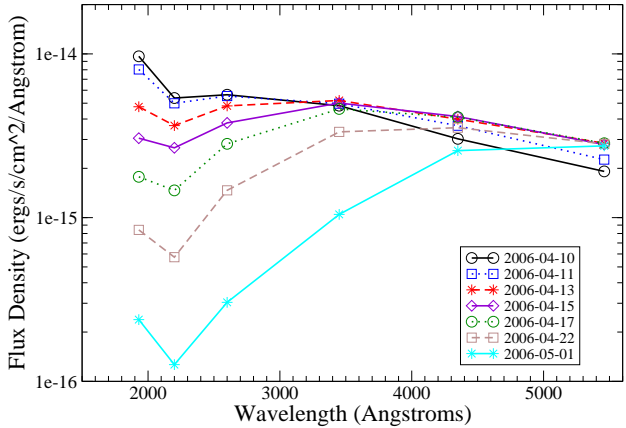,width=9.3cm,angle=-90,clip=}
	\end{picture}
	}
    \end{picture}
\caption{
Evolution of the spectral energy distribution (SED) of \sn\ in the
optical/UV wavelength band. The SED was produced from near-simultaneous UVOT
observations in all six filters obtained between days 1 and 26 after the 
explosion (from top to bottom).
} 
\label{fig5}
\end{figure}

\section{Summary}

Starting at an age of $\approx 1$~day after the outburst, our \S\ observations 
of \sn\ represent the earliest X-ray observation and detection of a SN to date,
apart from two explosions which were accompanied by a $\gamma$-ray signal
(as in the case of SN~1998bw/GRB 980425, which \B\ started observing 12~hrs after 
the outburst [Pian et~al.\ 2000] and SN~2006aj/GRB 060218, for which prompt X-ray 
emission was detected [Campana et~al.\ 2006]).
\sn\ faded below the \S\ XRT sensitivity limit within less than two weeks 
and a more sensitive \X\ observation recovered the SN three 
weeks after its explosion. Broad-band SEDs and in particular UV spectra of SNe
show the importance and give insights into line blanketing during the early phase
in the evolution.

Currently, \S\ is the only telescope capable of revealing the fast changes in the 
UV flux from SN explosions, offering sensitive constraints on the ionization state 
in the continuum and line formation region of the ejecta, the sources of metal 
line-blanketing. By extending the spectral coverage, this provides stronger 
constraints on the reddening, and complementary information, to what can be
deduced from optical observations alone. While there is ample evidence for diversity 
in type II SN optical spectra, \S\ observations can be used to address the existence 
of a corresponding diversity in the UV range. 

The detection of \sn\ in X-rays at such an early epoch, as well as the detection
of its fast optical/UV spectral evolution was made possible by the quick response 
of the \S\ satellite, and underlines the need for rapid observations of SNe in 
the UV and in X-rays. It should be pointed out that the response time of all 
previous X-ray observations of SNe (with missions such as \E, \A, \R, \C, and \X) 
were in the range between $>4$ days (SN~2002ap) and a few weeks. 
The bulk of all X-ray observations took place weeks to months after the explosions. 
Therefore, if \sn\ is more typical for a larger sample of core-collapse SNe, then 
the early production of X-rays would have been missed in each of the previous cases.
The response time of our ongoing \S\ observing program to study the prompt emission 
of SNe across the optical, UV, and X-rays is currently only limited by the time lag
between the explosion of a SN and its discovery at optical wavelengths and timely 
alert by the community.

\acknowledgments                                                               
We gratefully acknowledge support provided by STScI grant HST-GO-10182.75-A (P.A.M), 
NASA Chandra Postdoctoral Fellowship grant PF4-50035 (D.P.), and NSF grant AST-0307366 
(R.A.C). L.D acknowledges support for this work from the Scientific Discovery
through Advanced Computing (SciDAC) program of the DOE, grant DE-FC02-01ER41184 and 
from the NSF under grant AST-0504947. K.W.W. thanks the Office of Naval Research for the 
6.1 funding supporting this research. C.J.S. is a Cottrell Scholar of Research Corporation 
and work on this project has been supported by the NASA Wisconsin Space Grant Consortium. 
This work is sponsored at The Pennsylvania State University by NASA
contract NAS5-00136. We wish to thank N. Schartel and the \X\ SOC for approving and 
scheduling a \X\ DDT observation.
The research has made use of the NASA/IPAC Extragalactic Database (NED) which 
is operated by the Jet Propulsion Laboratory, California Institute of Technology, 
under contract with the National Aeronautics and Space Administration.

\end{document}